# Non-isothermal X-ray Emitting Gas in Clusters of Galaxies

Mark J. Henriksen

Department of Physics

University of North Dakota

Grand Forks, ND 58202-7129

and

Raymond E. White, III

Department of Physics & Astronomy

University of Alabama

Tuscaloosa, Alabama 35487

Abstract

We have analyzed X-ray spectra from six galaxy clusters which contain cooling flows: A85, A478, A1795, A2142, A2147, & A2199. The X-ray spectra were taken with the *HEAO1-A2* Medium and High Energy Detectors and the *Einstein* Solid State Spectrometer. For each cluster, we simultaneously fit the spectra from these three detectors with models incorporating one or more emission components comprised of either thermal or cooling flow models. Five of the clusters (all but A2142) are better fit by a multi-component model (a cooling flow plus one or two thermal components or a two thermal component model) than by isothermal models. In four of the clusters (A85, A1795, A2147, & A2199), we find evidence for cool gas outside of the canonical cooling flow region. These latter four clusters can be characterized by three temperature components: a temperature inversion in the central region, a hotter region with an emission-weighted temperature which is higher than that of an isothermal model fit to the entire cluster, and a cooler region with an emission-weighted temperature of $\sim 1$ keV. The cool component outside the cooling flow region has a large minimum emission measure which we attribute, in part, to diffuse cool gas in the outer cluster atmosphere. If at least some of the cool exterior gas is virialized, this would imply a radially decreasing temperature profile. Together with the density profiles we have found, this leads to a baryon fraction in gas which increases with radius and is larger than that for an isothermal cluster atmosphere. Consequently, if clusters of galaxies trace the mass distribution in the Universe, the gas mass fraction we have calculated for an isothermal gas (which is $\sim 15\%$) together with the nominal galaxy contribution ($\sim 5\%$) gives a baryon fraction of $\sim 20\%$. Using the upper limit to the baryon density derived from Big Bang nucleosynthesis gives a firm upper limit for $\Omega$ ($\sim 0.5$). The isothermal gas baryon fractions calculated here are lower than earlier estimates due to our utilizing a 3-component model together with data sets which allow us to remove the influence of the cooling flow on the integrated spectrum, as well as possible contaminating emission from unvirialized gas or discreet sources. *Subject headings* : galaxies:clustering-X-rays:galaxies-cosmology-dark matter

I. Introduction

The radial temperature distribution in the X-ray emitting gas in clusters of galaxies has been difficult to obtain in a model-independent manner using the available data. The primary reason is that most of the the spectral data taken prior to *ROSAT* and *ASCA* (i.e. from *Ginga*, *HEAO1-A2*, *Einstein* and *EXOSAT*) had no spatial resolution. These data have been used to derive isothermal temperatures for many clusters (Hatsukade 1989;

Mushotzky 1984; David et al. 1993; Edge et al. 1990). A few clusters have had temperature profiles derived with very coarse spatial resolution. For example, the Virgo cluster has had a rough temperature profile deduced from two pointings of the *Einstein* SSS (Lea, Mushotzky & Holt 1982), who found the center to be relatively cool) and a scanning observation with the *Ginga* LAC (Koyama, Takana, & Tawara 1991, who found the gas to be isothermal to the north and with a rising temperature to the south). Cowie, Henriksen, and Mushotzky (1987), Henriksen and Mushtozky (1986) and Henriksen (1987) fit a model utilizing a polytropic equation of state to the HEAO1-A2 data for eleven clusters which included Coma and Perseus. They and found evidence for non-isothermality in seven of the eleven clusters. The choice of a polytropic temperature and density relationship was justified because outside of the central region, gas processes are expected to proceed adiabatically since the radiative loss timescale is very long. The data used in their analysis was restricted to $> 2$ keV and reduces sensitivity to the cooling flow while still retaining sensitivity to a decrease in temperature with radius. In the Coma cluster (A1656), multiple $EXOSAT$ pointings were used by Hughes, Gorenstein & Fabricant (1988) and Edge (1990) to show that the the temperature declines beyond $25'$; spatially resolved spectra from the coded mask telescope *Spacelab 2* also show this decline, but with poor statistical significance (Watt et al. 1992). More recently, $BBXRT$ provided crude spatial information from a segmented detector construction; while it provided very good spectral coverage of the cooling flow regions of a few clusters, its small field of view relative to the total extent of cluster emission prevented it from mapping gas much outside their central regions with its limited number of pointings. Nonetheless, $ASCA$ observations of the Perseus (A426) (Arnaud et al. 1994) and $BBXRT$ observations of the Fornax (Serlemitsos et al. 1993) clusters show temperatures rising outward in the central region. The latest results from $ASCA$ observations of clusters are ambiguous in their implications regarding the temperature profile of the hot component of the intracluster medium. On the one hand, the inner region of nearby clusters do not appear to have temperature gradients (Mushotzky 1995), however, a number of those at higher redshift show steep temperature gradients (Markevitch 1995). These latter clusters may reflect a temperature profile peculiar to a recent post-merger system or they may simply show a temperature drop which occurs outside of the field of view for the nearby clusters.

Most of the pre-*ASCA* imaging data has come from the *Einstein* Imaging Proportional Counter (IPC) and the *ROSAT* Position Sensitive Proportional Counter (PSPC), which are relatively insensitive to temperature changes in the intracluster gas above ∼2 keV; PSPC observations require long integrations to detect potential temperature gradients. However, groups of galaxies have characteristic temperatures of ∼ 1 keV and therefore are ideally suited to the *ROSAT* PSPC for temperature profile determinations over large fractions of the groups' X-ray emission. In fact, the few groups with temperature profiles reported so far — HCG 62 (Ponman & Bertram 1993) and NGC 5044 (David et al. 1994) — show non-isothermal emission in the form of relatively cool gas both in the central regions and in the exterior regions of the intragroup medium. Deconvolutions of IPC data indicate the presence of relatively cool gas in the central regions of some clusters (Arnaud 1988). In other work, the *Einstein* Solid State Spectrometer (SSS) data from the central regions of clusters have been fit by a spectral model consisting of a cooling flow model plus an isothermal model in order to derive cooling flow accretion rates and column densities of neutral hydrogen (D. White et al. 1991). Large column densities of neutral hydrogen were found to be associated with the cooling flow clusters in this sample. *BBXRT* data (Mushotzky 1993) and a joint analysis of SSS and *Ginga* LAC data (R. White et al. 1994) has confirmed these results and placed new constraints on abundance profiles in cooling flow clusters.

One can obtain crude spatially resolved spectral information about clusters by combining spectra taken by instruments with different fields of view. In this paper, we present the results of such an analysis: we jointly fit *Einstein* SSS (small field of view) and *HEAO1-A2* (large field of view) spectral data for six cooling flow clusters to determine if there is cool intracluster gas in addition to the central cooling flows. Cool intracluster gas at large radius could potentially be in the form of virialized gas or unvirialized cool gas found in the outer parts of clusters in simulations of cluster formation and evolution (Thomas & Couchman 1991; Katz and White 1993).

Henriksen (1995) showed that the average baryonic mass fraction calculated for the HCG62, NGC5044, and NGC2300 groups is consistent with $\simeq 20\%$ when calculated out to the visible extent of the X-ray emission. This value is consistent with the canonical value for rich clusters. More data was recently obtained on N2300 and is consistent with a high baryon fraction, 10 - 16% (Davis et al. 1995). As with clusters, the value of the baryon fraction in groups depends on the extent of the gas. Unfortunately, this is typically poorly determined. If groups trace the mass density of the Universe, then this argues for a value of $\Omega \leq 0.30$. Pildis, Bregman, and Evrard (1995) report a similar baryon fraction in three additional groups.

II. Data

*A. Spectra*

We analyzed spectra taken with the *Einstein Observatory* SSS (Holt et al. 1979) and with the Medium and High Energy Detectors (MED and HED) on *HEAO1-A2* (Rothschild et al. 1979). A log of the spectral data can be found in Table 1.

The energy band of the SSS is 0.5–4.5 keV and the field of view is circular with a radius of 3 arc min. The combined MED and HED energy band covers 1–60 keV and the large field of view detectors of *A2* have a $3° \times 3°$ FWHM, pyramidal spatial response. The *A2* detectors have a well-determined internal background and the sky X-ray background is determined in a $6°$ offset mode. The energy resolution for the *A2* detectors is 15 - 20% at 6 keV. The *A2* has a systematic calibration error of less than 1% (Marshall et al. 1979), which is not included in the modeling because it is negligible when compared to the random error of even the highest signal-to-noise *A2* cluster observation, Coma. The highest signal-to-noise in any channel in the *A2* observation of Coma is 2%. The observations presented in this paper are of significantly poorer quality than this and do not warrant adding in the systematic calibration error.

The background subtracted spectral data for each detector are grouped, if necessary, to insure that there are enough counts ($\sim$20–30) in each channel to give accurate parameters in a $\chi^2$ test (Nousek and Shue, 1989), without degrading the energy resolution (approximately 3 channels),

Since each cluster has multiple SSS spectra, we tested whether their normalizations are equal by considering both tied and independently varying normalizations when using a fiducial isothermal model. Pointing error or drift during integrations could result in variations in the incident flux on the relatively small aperture of the SSS. In fitting isothermal Raymond-Smith models to the SSS data, we found for three of the clusters (A2147, A2199, & A478) that letting the normalizations vary reduced $\chi^2$ significantly (with >90% confidence); this reduction in $\chi^2$ is of the order of that seen when we add an additional abundance or absorption component to the multicomponent models we consider below. The best-fit values of the independently varying normalizations were all within their respective 90% confidence ranges of one another, however. Joint fitting of the models, which typically have 8 free, physically interesting parameters, makes it impractical to let all of the normalizations vary for each spectrum since this would add 3 to 6 more free parameters. Typically, 40–60% of the total observation time is contained in one SSS spectrum, so we have used the single SSS spectrum with the longest observation time for each cluster to make a less ambiguous determination of the best fit model required for each cluster. However, to insure that we have not overlooked a subtle absorption or abundance component in the SSS and to minimize the error bars on interesting paprameters, we refit the best fitting model to all of the SSS spectra (with tied norms) plus the *A2* spectra.

Some of the SSS observations are characterized by periods of ice formation on the detector which increases the absorption of soft X-rays. The ice formation problem has been modeled (Christian et. al 1992) and observations with high ice formation during an integration (i.e. those with changes in the ice parameter of $> 0.1$) are not used.

The pointing of the SSS for each observation is checked against the centroid of the IPC emission from the cluster for coincidence. There are multiple observations for most clusters and short observations, 1–2 ksec, are not used when they are a small fraction of the total observing time in order to minimize any uncertainties associated with requiring the total fluxes to be equal (tied norms) in the model fitting procedure.

Source contamination must be considered in data from large collimators such as the large field of view detectors on *HEAO1-A2*. The possibility that the cool components reported in this paper are due to contaminating sources was checked using *ROSAT* PSPC fields. For A2199, A2147, A85, and A1795, the contribution from non-cluster sources, normalized to the *A2* energy band is 5%, 7%, 14%, and 20% of the measured cool component emission. The PSPC field of view has a one degree radius. The *A2* HWHM is 1.5° so that it extends beyond the PSPC field of view. Thus, an estimate of the contribution from non-cluster sources to the cool component could be slightly higher. The conversion of PSPC flux to the *A2* bandpass uses the best fit spectrum of non-cluster sources to estimate their contribution. The best fit spectra are not, in general, similar to that found for the cool component.

*B. Images*

The analysis of the images and results for five of the six clusters is taken from Henriksen (1987). The procedure for preparing the Einstein images is similar to that described in Jones and Forman (1984). The most significant difference is the treatment of the central region of the cluster surface brightness profile. We fit a model consisting of the profile, $S/S_c = (1 + (\frac{r}{a})^2)^{-2\delta + 1/2}$ with a Gaussian function, $I_c \exp(-\frac{1}{2}(\frac{r}{\sigma})^2)$, which makes a significant contribution in the central region. In all of the clusters, the single profile could not fit the central region, however, the addition of the Gaussian provided a good fit to all of the data. The total emission in the Gaussian component is $2\pi\sigma^2 I_c$ and is given in Table 7 along with the other fit parameters, $\delta$ and a. The central density is derived from the normalization of the HEAO1-A2 spectrum in the isothermal fit, $\frac{1}{4\pi D^2} \int n_e^2 dV$ cm$^{-5}$.

### III. Spectral Models

We fit spectral models ranging from the simplest (a single Raymond-Smith thermal component) to complex (cooling flow plus two thermal components). The single thermal spectral model is characterized by five parameters: the temperature; the metal abundance (specified as a fraction of Solar); the column density of neutral hydrogen in the line of sight $N_H$, the redshift; and the normalization. Given the spectral resolution of this data, the redshift is fixed at that determined optically, leaving four free parameters.

The predicted spectrum for the thermal components is a product of the cooling function $\Lambda(T)$ (in units of ergs cm$^3$ sec$^{-1}$), the emission measure $\int n_e n_i dV$ (in units of cm$^{-3}$), and the absorption, exp($-\sigma N_H$), folded through the energy response matrix of the detector. The cooling function for a low density plasma in collisional ionization equilibrium is described in Raymond & Smith (1976) and the absorption cross-sections, $\sigma$, is given in Morrison & McCammon (1983). The column density is fixed at the galactic value (Stark et al. 1992), although the hypothesis of additional absorption in the cooling flow region is tested. The *A2* spectra are not sensitive to the column density, but the SSS spectra are; if there is additional absorption above the galactic value then fixing the column density at the Galactic value will raise the temperature and lower the abundance of the gas in spectral fits to the SSS data.

The cooling flow spectral model (CFRS) is an empirical model Mushotzky & Szymkowiak (1988) which allows measurement of the mass inflow rate in the cooling flow. There are potentially 5 free parameters in this model: the mass accretion rate, the temperature from which the gas cools ($T_{high}$ in Table 4), the temperature to which the gas cools, the elemental abundance relative to Solar abundances, and the slope of the power-law which describes the distribution of emission measure versus temperature. The temperature which the gas cools to is fixed at 80 eV so that there are four free parameters. The temperature the gas cools from is tied to the temperature of the hot thermal component which dominates the *HEAO1-A2* spectrum. It is not necessarily equal to the emission-weighted temperature of the entire cluster derived from isothermal models (as assumed in D. White 1991). The abundance in the cooling flow is tied to the abundance of the *HEAO1-A2* spectrum, which is dominated by a strong Fe K$\alpha$ feature at 6.7 keV. Initially, the abundance is required to be constant across the cluster. The slope is left as a free parameter. The best fitting model is also refit with the slope fixed at 0 (corresponding to isobaric cooling). Freeing the slope, evaluated by the *F*-test, does not improve the fit to the data for A478, A2199, and A1795. For A85, there is a 75% probability that the fit is improved by freeing the slope and there is a 90% chance that freeing the slope improves the fit for A2147. As discussed earlier, one of the assumptions in fitting data sets such as the SSS is that all of the pointings have the same flux. For comparison, letting the relative flux in the SSS pointings vary for A2147 produces as much of an improvement as freeing the slope in the cooling flow model. The cooling flow mass inflow rates quoted in this paper utilize a model with a variable slope. Specific values of the slope describe different dynamical states of the cooling flow (e.g. radial pressure gradients) and with higher quality data, this parameter could be determined. A second Raymond & Smith (RS) component is added to the cooling flow plus a RS (CF2RS) and is required to contribute only to the *HEAO1-A2* spectra to directly detect cool gas in the outer cluster atmosphere.

## IV. Results

Table 2 summarizes the models fit to the three spectral data sets (SSS, MED, and HED). The best-fit value of $\chi^2$ and the number of degrees of freedom are shown so that the significance of adding more degrees of freedom in the form of model parameters is apparent. The models are arranged so that the number of model parameters increases across the table. Fits utilizing all of the SSS data files are shown in parenthesis, while those which use the highest signal-to-noise observation are not.

$\chi^2$ and the degrees of freedom in Table 2 show that the isothermal model is a poor fit to all of the cluster spectra except A2142. For comparison to the spectral results from the Einstein, EXOSAT, and GINGA, we present the best fitting isothermal parameters in Table 3. An additional Raymond & Smith thermal component is added to the previous isothermal model, creating a dual-temperature model denoted as 2RS. Fixing the redshift and tying the abundance of this second component to the first leaves an additional two free parameters: an independently varying temperature and normalization (emission measure). Five of the clusters (all but A2142) are better fit by a multi-component model (2RS, CFRS, or CF2RS) than an isothermal. Table 4 contains the best fit parameters with 90% error bars for the preferred model for each cluster. The preferred model is the best fitting model exclusive of the two Raymond and Smith component model. Two component models have generally been attributed to a spectrum consisting of a cooling flow and the hot atmosphere. Since our purpose is to search for cool gas in the outer atmosphere, we merely use the cooling flow model to fit the spectrum in the central region of the atmosphere. We note that for A478 and A1795, the two component model is as good of a fit to the data as the best fitting cooling flow model. This area is investigated further by Henriksen and Silk (1995) who offer an alternative interpretation of the two component spectrum: that the cool component may be due to bright early type galaxies or the intergalactic gas of accreted groups. In this interpretation, the two component model is an accurate description of the gas. For two of the clusters (A2142, A2199), use of all of the SSS data files reduced the error bars and is used in Table 4, otherwise, the values are for the single SSS file with the longest exposure time (since this avoids the variable normalization problem). The mass inflow rates rates are consistent with those found by White et al. (1991) though these authors used only the SSS data with $T_{high}$ fixed at the isothermal temperature of the cluster determined by Edge (1989).

In order to determine whether there is cool gas outside of the cooling flow, we added a temperature component to the CFRS model and constrained it to be in the *HEAO1-A2* field of view, but outside the SSS field of view. Initially, this model had one more free parameter than the CFRS model since the normalization of the cool component is required to be equal to the hot component. This test was done using a reduced number of SSS data sets to remove the uncertainty due to flux variation in the SSS pointings. There are 8 free parameters and *initially* the hot and cool components normalizations were tied inorder to introduce only one new free parameter and look for changes in Chi-square. Later, the normalizations were untied; this additional degree of freedom is not significant. The $F$-test shows that there is a cool component in 4 of 6 clusters with greater than 99% confidence. Figure 1 shows the CFRS model (no outer component) fit to A1795. Comparison to Figure 2, with cool component added shows that the part of the spectrum with requires the cool gas is spread over primarily 8 grouped channels in the energy range of 1.5 - 2.5 keV range of the Medium Energy Detector. This together with the fact that two out of six clusters do not require the additional cool gas argues strongly against either systematic errors in calibration or a flakey channel as the source of the cool gas.

However, the cool component flux evaluation was done using all of the data sets and allowing the normalization of the cool component to be free. The normalization is a fit parameter derived from fitting the model to the spectral data. Table 5 contains the 90% error on the normalization or emission measure for the cool component outside of the central region. The normalization (in units of $10^{14}$ cm$^{-5}$) is related to the flux by the equation, Flux = $(Norm) \times (K - correction) \times (\Lambda(T))$. The normalization is the emission measure and is given by $\frac{1}{4\pi D^2} \int n_e^2 dV$, where D is the distance to the cluster, n$_e$ is the electron density, and $\Lambda$(T) is the cooling function. We find that the cool component contributes 16–47% of the 1–10 keV flux in the *A2* detector. Taking the minimum (90% confidence) emission integral for the cool component gives a contribution of 5, 19, 20, 31% for A85, A2147, A2199, and A1795, respectively, to the total flux. The flux of non-cluster sources in the same energy band to the cool component, in the same order, is 14%, 7%, 5%, and 20%. It is interesting to note that there is no correlation between the flux of the cool component and the flux of the non-cluster sources. Ordering the clusters from lowest cool component flux to highest is: A85, A2147, A1795, and A2199. Ordering by non-cluster source flux gives: A2147, A85, A2199, and A1795. We also fit a cooling flow added to two RS components with the constraint that all of the cool material (cooling flow plus cool RS component) be in the SSS field of view; this is a way to test the robustness of the cooling flow model and whether there is additional cool material associated with the cooling flow as opposed to the additional cool gas originating outside of the central region. For all of the clusters which require a a cool component, the restriction that it be contained in the central region provides a significantly worse fit than if it is all outside of the central region.

The best fitting model for each cluster is used to test whether or not the data require additional absorbing material or increased abundance in the cooling flow. A single parameter is set free (either the absorption or the abundance in the cooling flow) and the data are refit. $\chi^2$ is improved for both free abundance and absorption in one cluster, A2199. The probability that the change in $\chi^2$ is significant based on the *F*-test is 95% for a higher abundance in the cooling flow component, and 90% for higher absorption in the cooling flow. Two other clusters are better fit with a higher column density than Galactic: A478, and A2142. A478 has a higher absorption in the cooling flow with greater than 99% confidence. The mass inflow rate with Galactic column density is 657 $M_\odot$ yr$^{-1}$ for A478 and 1275 $M_\odot$ yr$^{-1}$ with additional absorption in the cooling flow; the approximate best fit intrinsic absorption is 1.4 $\times 10^{21}$ cm$^{-2}$, nearly equal to the Galactic value. The best fit intrinsic column density for A2142 is 7.$\times 10^{20}$ cm$^{-2}$, approximately twice the Galactic value and it is significant at $> 99\%$ confidence. Refitting using all of the SSS data with normalizations tied indicates an absorption and abundance gradient in A1795 with the cooling flow having a lower abundance than the ambient components. Comparison of the abundance from fitting an isothermal model to the flux weighted abundance calculated from the best fit abundances of the individual components indicates that the abundances is likely too high in the abundance separation found with the multicomponent models. We fit only the SSS files for A1795 with all parameters tied. We then allowed all of the column densities to be different or all of the abundances to be different or all of the temperatures to be different or all of the normalizations to be different, in turn. $\chi^2$ drops by the same amount for either free norms or free column densities so that the increased absorption in A1795 is a tentative result. Letting the column density and abundance go free in the cooling flow indicates that they are larger than in the hot component with 80% probability, however, the 1$\sigma$ error is consistent with a constant value.

V. Discussion

*A. Non-isothermal Gas*

By fitting a range of models, from single to multiple components, we have found that the best fitting model for the central region consists of either a 2RS or a CFRS. There is cool gas outside of the central 3 arcmin ( 0.2–0.5 Mpc). The gas has a large emission measure and we argue below that it can not be from cooling flow gas extending beyond the *SSS* field of view. Because the cool component does have large emission measure, we conclude it must occupy a large volume in the outer atmosphere of the cluster. PSPC observations by Briel, Henry, & Bohringer (1991) have confirmed that the X-ray emitting gas may extend out to 3 Mpc in clusters (Henriksen & Mushotzky 1985). The *A2* detectors have a significantly larger field of view than *Ginga* or *EXOSAT* and are more likely to contain a contribution from cool gas in the outer atmosphere since the density of the gas typically decreases as r$^{-2}$, and there must be a large contributing volume to be detectable in the cluster spectrum. Indeed, the contribution to the flux in the *HEAO1-A2* detectors indicate that the cool component is 16%, 23%, 40%, and 47% of the total flux in the 1–10 keV band for A85, A1795, A2147, and A2199, respectively. The cooling flow component is 47%, 36%, 9%, and 13% for these clusters, respectively. The cool component can not be attributed to poor modeling of the cooling flow component since it is always substantial and in half the clusters greater than the cooling flow component. Indeed one does not see an obvious relationship between $\dot{M}$ and the cool component such that it would be attributed to poor modeling of the cooling flow. In the case of A478, which has the largest $\dot{M}$, no cool component is even found. One should keep in mind that the cooling flow is also constrained by the $SSS$ data. A related possibility is that the cool component is due to cooling flow gas spilling out of the $SSS$ field of view. Cooling flow temperature profiles generally increase with radius. If this is the case, then the part of the cooling flow detected as the cool component by the $A2$ detector would always come from gas located nearer to the hot component than that part of the cooling flow detected by the $SSS$. This would imply that the temperature of the cool component should be substantially higher than the cooling flow component. Since this trend is not found, it is unlikely that the cool component is due to cooling flow gas outside of the $SSS$ field of view.

B. Mass Components

X-ray spectral and imaging observations can be used to calculate the gas mass and total gravitating mass for groups and clusters. This was done recently by David et al. (1994) using ROSAT imaging observations. What we have done differently is performing a spectral deconvolution of the temperature components in the cluster gas. We have shown that there are at least three temperature components which affect the spectrum and combine to give the emission weighted temperature of a cluster measured by a large field of view detector such as $HEAO1-A2$, $GINGA$, $EXOSAT$, or $MPC$. The cooling flow must be removed since only the gas which has not suffered substantial radiative losses is strictly in hydrostatic equilibrium with the cluster potential. The low temperature gas in the diffuse cluster atmosphere which we have found may be due to a decrease in the cluster gas temperature with radius, similar to what is observed in groups, but it may also have other contributions including unvirialized gas. Even if the hot atmosphere is isothermal, removal of the cool components results in a more accurate hot component temprature and therefore a more accurate total gravitating mass calculation. To parametrize the temperature profile, we use a polytropic temperature and density relationship which is a convenient empirical relationship. The temperature profile is,

$$T(r) = T_c[1 + (\frac{r}{a})^2]^{-\delta(\gamma-1)}$$

and the density profile is given by,

$$n(r) = n_c[1 + (r/a)^2]^{-\delta}.$$

Delta is determined from fitting the surface brightness radial profile, if the spectral response of the detector is insensitive to the temperature changes in the gas or if the gas is isothermal. Because the EINSTEIN IPC is generally insensitive to temperature changes in the gas one can get a well determined density profile for clusters. For the clusters analyzed in this paper, $\delta$ is determined to 5 - 10%. The density profile is integrated, assuming spherical symmetry, and the gas mass is given by,

$$M_{gas}(<x_{cl}) = 2.8 \times 10^{13} (\frac{a}{0.25 Mpc})^3 (\frac{n_c}{5 \times 10^{-3}}) \times \int_0^{xcl} \frac{x^2 dx}{(1+x^2)^{\delta}} \; h_{50}^{-5/2} \; M_\odot$$

The gas mass is calculated using the core radii, $\delta$, and central densities in Table 6 and is contained in Table 7. The cluster masses are calculated out to 10 core radii. The extent of the gas was derived from a comparison of the Einstein IPC and HEAO1-A2 fluxes for these clusters (Henriksen and Mushotzky 1985) and has been confirmed by Mushotzky (1994) using PSPC images of A1795 and A478.

The temperature and density profiles are substituted into the equation of hydrostatic equilibrium to derive the total mass distribution:

$$M_{tot} = 1.92 \times 10^{14} \; \delta\gamma(\frac{T_c}{10 keV})(\frac{a}{0.25 Mpc}) \; \frac{x^3}{(1+x^2)^{(1+\delta(\gamma-1))}} h_{50}^{-1} \; M_\odot$$

The ratio of the gas to total mass gives the baryon fraction in the gas; this is calculated over the same extent as the gas and shown in Table 8. Henriksen (1994) found that the lower limit to the baryon fraction ($\sim 20\%$) in groups using observations reported by Henriksen & Mamon (1994), David et al. (1994), Mulcahey et al. (1993), and Ponman & Bertram (1993) simply by using the visible extent of the gas as to calculate the baryon fraction.

The baryon fraction is calculated for the extent of the gas in the PSPC observations and is a lower limit since it is increasing for the NGC2300, NGC5044, and HCG 62 groups. The extent of the gas is, in general, poorly determined. Henriksen & Mamon (1993) shows that in gases where the temperature profile decreases with radius, the gas mass will increases faster than the total mass and the baryon fraction will increase with radius. Davis et al. (1995) have obtained more PSPC data for the N2300 group and have revised the baryon fraction found by Mulchaey et al. (1993) to the range 10 - 16%. The baryon fraction in a non-isothermal distribution characterized by $\gamma$ relative to an isothermal distribution is given by:

$$\frac{f_\gamma}{f_{iso}} = \gamma \left(1 + \left(\frac{r}{a}\right)^2\right)^{\delta(1-\gamma)}$$

For these clusters, if the temperature profile is characterized by a $\gamma$ of $\geq 1.1$ (as favored in the simulations of Thomas & Couchman) in which the cool gas is due in part to virialized gas and also unvirialized cold gas, then the average baryon fraction in gas is 23% (see Table 8). If all of the cool gas found outside of the cooling flow in this paper were virialized, this would imply a larger $\gamma$ in which the baryon fraction may approach 1 at large radii (i.e., the visible baryons in gas and galaxies approaches the total mass of the cluster). The firm lower limit is for the isothermal atmosphere with the hot gas tempertures in Table 4. Using these values, the baryon fraction in gas is $\geq 15\%$. With an additional 5% baryons contributed by galaxies, the lower limit would be 20%. David, Jones, and Forman (1995) confirm our results using an isothermal model applied to observations of groups and clusters. The presence of cool gas in the cluster atmosphere simply lowers the measured isothermal temperature and gives a smaller value for the total cluster mass and a larger gas mass fraction. Two of our clusters, A85 and A1795 are in their sample. The isothermal temperatures for these clusters in Table 3 agree very well with their isothermal temperature from $GINGA$ and $EXOSAT$. The difference in spectral modelling accounts for the slightly higher value of the gas mass fraction they find.

## Conclusions

A radial temperature profile is necessary to derive the gravitating mass component depen-dence on radius. We have shown that there are departures from isothermality for a number of rich clusters, in addition to that associated with the cooling flow, and argued that it is in the outer cluster atmosphere. A temperature profile which decreases with radius results in an increasing baryon fraction with radius. This implies that the baryon fraction may be even higher than 20% and implies a low $\Omega$ ($< 0.3$) Universe. If we consider the uncertainty in $\Omega_b$ measured from Big Bang Nucleosynthesis, then using the upper limit on $\Omega_b$ of 0.1 (Krause 1994) along with the lower limit of the baryon fraction in clusters of 0.2, we derive a very *robust* upper limit on $\Omega$ for the Universe of $0.5 h_{50}^{-1/2}$. The primary assumption here is that clusters adequately trace the mass in the Universe.

## References

Arnaud, K.A. 1988, In *Cooling Flows in Clusters of Galaxies*, 31.

Arnaud, K.A., et al. 1994, *ApJ*, 436, L67.

Bolte, M., 1994, BAAS, 24, 1397.

Briel, U.G., Henry, J.P., and Bohringher, H. 1992, *AA*, 259, L31.

Christian, D.J., Swank, J.H., Szymkowiak, A.E., and White, N.E., 1992, *Legacy*, 1, 38.

Cowie, L.L., Henriksen, M.J., and Mushotzky, R.F., 1987, *ApJ*, 317, 593.

David, L.P., Jones, C., and Forman, W., 1995 (preprint).

David L.P., Forman, W., and Jones, C., and Daines, S., 1994, *ApJ*, 428, 544.

David, L., Slyz, A., Jones, C., Forman, W., Vritilek, S., and Arnaud, K. 1993, *ApJ*, 412, 479.

de Vaucouleurs, G., 1993, *ApJ*, 415, 10.

Edge, A.C., 1990, Ph.D. Thesis, University of Leicester.

Edge, A., Stewart, G., Fabian, A., and Arnaud, K. 1990, *MNRAS*, 245, 559.

Eyles, C.J., Watt, M.P., Bertram, D., Church, M.J., Ponman, T.J., Skinner, G.K., Willmore, A.P., 1991, *ApJ*, 376, 23.

Hatsukade, X. 1991 Ph.D. Thesis, University of Osaka.

Henriksen, M.J., and Mushotzky, R.F. 1985, *ApJ*, 292, 422.

Henriksen, M.J., and Mushotzky, R.F. 1986, *ApJ*, 302, 287.

Henriksen, M.J., 1987 Ph.D. Thesis, University of Maryland.

Henriksen, M.J., and Mamon, G.A. 1993, *ApJ(Letters)*, **421**, L63.

Henriksen, M.J., 1995, in *Critique of Sources of Dark Matter in the Universe*, World Scientific press.

Henriksen, M., and Silk, J., 1995, *PASP*, in press.

Holt, S.S., et al., 1979, *ApJ*, 234, L65.

Hughes, J., Gorenstein, P., and Fabricant, D., 1988, *ApJ*, 329, 82.

Jones, C., and Forman, W., 1984, *ApJ*, 276, 38.

Katz, N., and White, S.D.M., 1993, *ApJ*, 412, 455.

Koyama, K., Takano, S., and Tawara, Y., 1991, *Nature*, 350, 135.

Krause, L. 1994, in *Critique of Sources of Dark Matter in the Universe*, World Scientific press.

Lea, S.M., Mushotzky, R., and Holt, S.S., 1982, *ApJ*, 262, 24.

Lowenstein, M., 1994, *ApJ*, 431, 91.

Marshall, F.E., et al. 1979, *ApJ*, 235, 4.

Morrison, R., and McCammon, D. 1983, *ApJ*, 270, 119.

Mushotzky, R.F., 1984, *Physica Scripta*, T7, 157.

Mushotzky R.F., and Szymkowiak, A.K. 1988, In *Cooling Flows in Clusters of Galaxies*.

Mushotzky, R.F. 1992 preprint.

Mushotzky, R.F., 1994, Paper presented at Winter AAS Meeting.

Nousek, J.A., and Shue, D.R., 1989, *ApJ*, 342, 1207.

Petre, R., 1993, *ApJ*, 413, 518.

Pildis, R., Bregman, J.N., and Evrard, A., 1995, *ApJ*, 443, 514.

Ponman T., and Bertram, D., 1993, *Nature*, 363, 51.

Raymond, J. and Smith, B.W., 1977, *ApJ. Supplement*, 35, 419.

Serlemitsos, P.J., Lowenstein, M., Mushotzky, R.F., Marshall, F.E., and Petre, R., 1993, *ApJ*, 413, 518.

Stark, A., et al. 1992, *ApJ Supplement*, 79, 77.

Steigman, G. 1989, in Cosmic Abundances of Matter, ed. C. Waddington (Minneapolis: AIP), 310.

Stewart, G.C., Fabian, A.C., Forman, W., and Jones, C. 1984, *ApJ*, 285, 1.

Watt, M.P. et al. 1992, *MNRAS*, 258, 738.

White, D. A., Fabian, A.C., Johnstone, R.M., Mushotzky, R.F., Arnaud, K.A., 1991, *MNRAS*, **252**, 72.

White, R.E. III, Day, C., Hatsukade, I., Hughes, J.P., 1994, *ApJ*, 433, 583.

Figure Captions

Figures 1: Contributions to $\chi^2$ for the isothermal + cooling flow model are shown for the A1795 cluster. The MED is shown which has greater sensitivity at lower energies and requires the additional cool component. Comparison to Figure 2 shows that the cool component does not effect a single channel but effects at least 8 channels between 1.5 - 2.5 keV.

Figure 2: Contributions to $\chi^2$ versus energy are shown for the A1795 cluster. The model used here is the same as in Figure 1 with the addition of a thermal component in the A2 field of view, but not in the central 3 arc min. Comparison to figure 2 shows that the effect of the second thermal component is to reduce $\chi^2$ by 17 for 2 additional degrees of freedom to give a reduced $\chi^2$ of 1.

| TABLE 1: Observation Characteristics | | | | |
|---|---|---|---|---|
| Cluster | Z | SSS Obs length (sec) | A2 MED Obs length (sec) | A2 HED obs length |
| A85 | 0.0518 | 1433<br>2662<br>6512 | 5933. | 7361. |
| A478 | 0.090 | 4096.<br>3236. | 1712. | 2606. |
| A1795 | 0.0621 | 7578.<br>5079.<br>5571. | 10355. | 12188. |
| A2142 | 0.090 | 14213<br>5243<br>4587 | 2455. | 3926. |
| A2147 | 0.036 | 5242.<br>6226.<br>7250. | 2162. | 3954. |
| A2199 | 0.031 | 8069.<br>5160.<br>9420. | 3423. | 3911. |

| Cluster | 1 RS | 2 RS | 1 RS + CFLOW | 2 RS + CFLOW |
|---------|------|------|--------------|--------------|
| A85 | 192/181 (333/314) | 184/178 | 182/178 | 169/177 (308/309) |
| A478 | 196/120 (218/139) | 129/118 | 143/117 (175/134) | - |
| A1795 | 194/141 (364/271) | 164/139 | 171/139 | 165/138 (309/266) |
| A2142 | 179/184 | - | - | - |
| A2147 | 181/154 (359/247) | 170/152 | 169/152$^2$ | 163/151 (342/242) |
| A2199 | 119/118 (319/265) | 111/116 | 113/116 | 106/115 (285/260) |

TABLE 2: [1]Basic Models: $\chi^2$/Degrees of Freedom

Fits using all of the SSS files shown in parenthesis

| Cluster | kT (keV) | Abundance |
|---------|----------|-----------|
| A85     | 5.1 - 6.2 | 0.24 - 0.58 |
| A478    | 3.4 - 4.4 | 0. - 0.25 |
| A1795   | 5.2 - 6.0 | 0.41 - 0.73 |
| A2142   | 8.5 - 11.0 | 0.23 - 0.86 |
| A2147   | 3.9 - 6.0 | 0.12 - 1.23 |
| A2199   | 3.4 - 3.8 | 0.57 - 0.84 |

TABLE 3: Temperatures and Abundances from Isothermal model

| Cluster | kT$_{high}$ | kT$_{low}$ | Abundance | $\dot{M}$ |
|---|---|---|---|---|
| | | TABLE 4: Best Fit Model Parameters | | |
| A85 | $8.6^{+3.4}_{-1.7}$ | $1.0^{+0.39}_{-0.28}$ | $0.46^{+0.39}_{-0.20}$ | $148^{+72}_{-93}$ |
| A478 | $8.2^{+3.4}_{-1.7}$ | - | $0.^{0.91}_{0.20}$ | $1275^{2225}_{795}$ |
| A1795 | $9.0^{+1.5}_{-1.5}$ | $0.68^{+0.15}_{-0.18}$ | $0.68^{+0.27}_{-0.22}$ | $220^{+40}_{-80}$ |
| A2142 | $9.5^{+1.5}_{-1.5}$ | - | $0.44^{+.29}_{-.44}$ | - |
| A2147 | $8.0^{+9.0}_{-3.0}$ | $1.25^{1.25}_{-0.62}$ | $1.4^{+2.4}_{-0.92}$ | $22^{+11}_{-14}$ |
| A2199 | $4.3^{+1.2}_{-0.5}$ | $0.64^{+.36}_{-.34}$ | $0.73^{+.12}_{-.22}$ | $99^{+60}_{-47}$ |

TABLE 5: $10^{-14} \times$ Emission Integral for Cool Component

| Cluster | Emission Integral |
|---------|-------------------|
| A85     | 0.0175 - 0.0525   |
| A1795   | 0.033 - $\infty$  |
| A2147   | 0.0074 - 0.24     |
| A2199   | 0.035 - 68.       |

| TABLE 6: Best Fit Model Parameters | | | | |
|---|---|---|---|---|
| Cluster | $\delta$ | a (mpc) | $n_c \times 10^{-3}$ cm$^{-3}$ | $L_{Gauss}$ $10^{43}$ ergs sec$^{-1}$ |
| A85 | 1.10 - 1.19 | 0.42 - 0.46 | 2.8 - 3.7 | 2.3 - 2.5 |
| A478 | 1.05 - 1.43 | 0.31 - 0.63 | 8.9 - 2.8 | 2.8 - 5.2 |
| A1795 | 0.96 - 0.92 | 0.21 - 0.18 | 6.2 - 4.1 | 1.6 - 2.1 |
| A2142 | 1.43 - 1.80 | 0.71 - 1.00 | < 2. | 4. - 7.2 |
| A2147[a] | < 0.68 | 0.17 - 0.31 | 1.6 | |
| A2199 | 1.06 - 1.19 | 0.22 - 0.38 | 1.8 - 3.6 | 0.6 - 0.7 |

[a] Magri et al. 1988

TABLE 7: Gas Mass Fractions in Clusters

| Cluster | $M_g$ ($10^{14} M_\odot$) | $M_{iso}$ ($10^{15} M_\odot$) | $f_{iso}$ | $f_{\gamma=1.1}$ | $f_{\gamma=1.2}$ | $f_{\gamma=1.3}$ |
|---------|---------------------------|-------------------------------|-----------|------------------|------------------|------------------|
| A85 | $5.3^{+0.8}_{-0.7}$ | $3.3^{+1.0}_{-0.7}$ | $0.16^{+0.06}_{-0.07}$ | $0.25^{+0.09}_{-.11}$ | $0.39^{+.14}_{-.18}$ | $0.61^{.24}_{-.29}$ |
| A478 | $8.8^{+10.2}_{-6.6}$ | $3.6^{+0.8}_{-0.8}$ | $0.24^{+.19}_{-.16}$ | $0.39^{+.33}_{-.28}$ | $0.64^{+.36}_{-.47}$ | $1.0_{-.75}$ |
| A1795 | $1.4^{+0.5}_{-0.5}$ | $1.2^{+.11}_{-.10}$ | $0.11^{+0.06}_{-0.04}$ | $0.16^{+0.09}_{-0.06}$ | $0.23^{+.12}_{-0.09}$ | $0.32^{0.18}_{-0.12}$ |
| A2142 | $6.7^{+0.5}_{-0.6}$ | $9.8^{+1.3}_{-1.4}$ | $0.07$ | $0.13^{+0.1}_{-0.0}$ | $0.25^{+0.04}_{-0.02}$ | $0.49^{+0.13}_{-0.08}$ |
| A2147 | $2.0^{+2.3}_{-0.5}$ | $9.8^{+5.2}_{-1.9}$ | $0.20^{+0.34}_{-0.15}$ | $0.25^{+0.42}_{-0.19}$ | $0.31^{+0.54}_{-0.24}$ | $0.40^{+0.60}_{-0.31}$ |
| A2199 | $1.4^{+1.9}_{-0.9}$ | $1.1^{+0.2}_{-0.1}$ | $0.13^{+0.13}_{-0.12}$ | $0.20^{+.20}_{-0.12}$ | $0.31^{+.33}_{-.22}$ | $0.48^{+0.52}_{-0.31}$ |

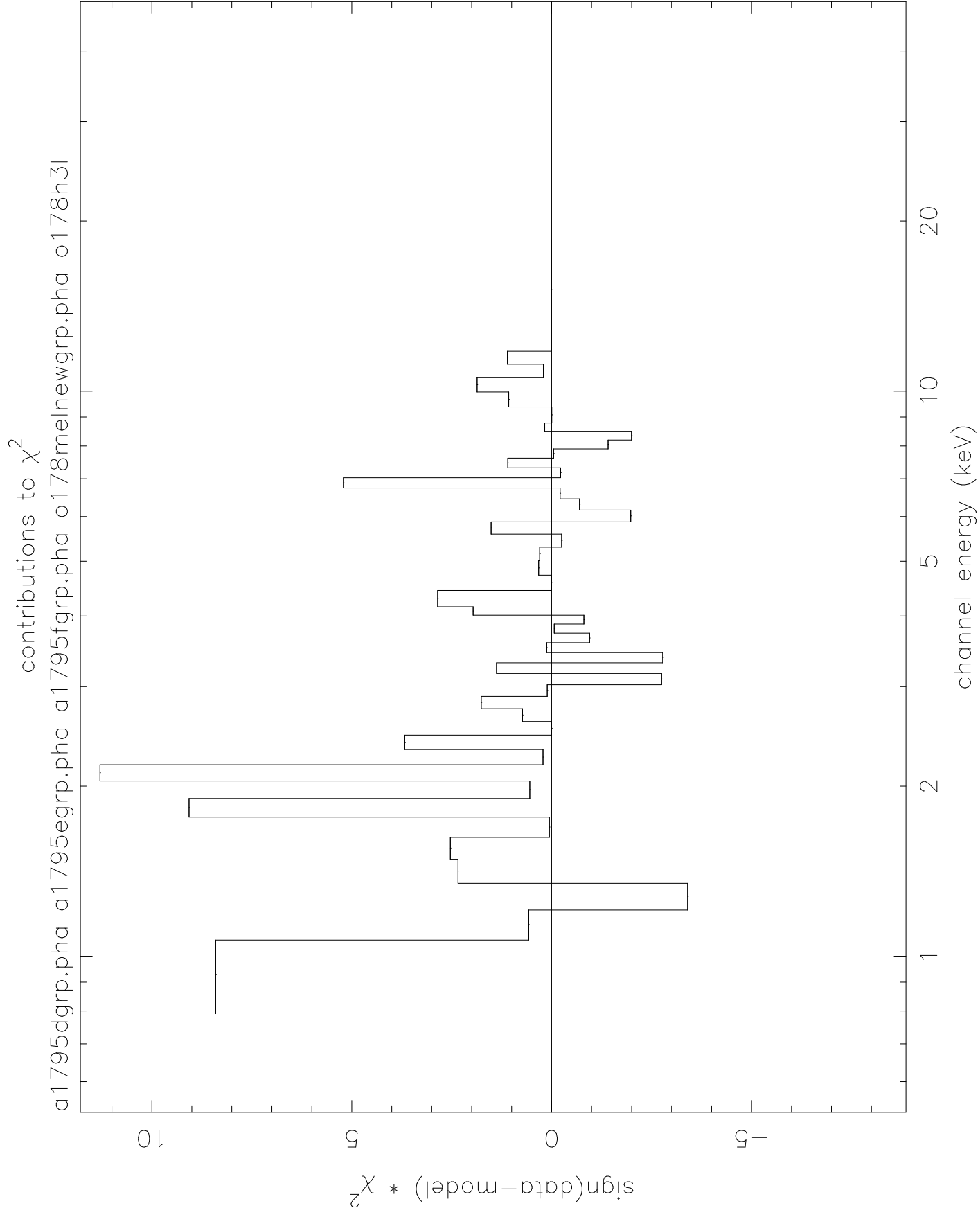

# XSPEC

For help : xanprob@athena.gsfc.nasa.gov

On our mailing list ? If not then send the message subscribe XANADU Your Name to listserv@athena.gsfc.nasa.gov

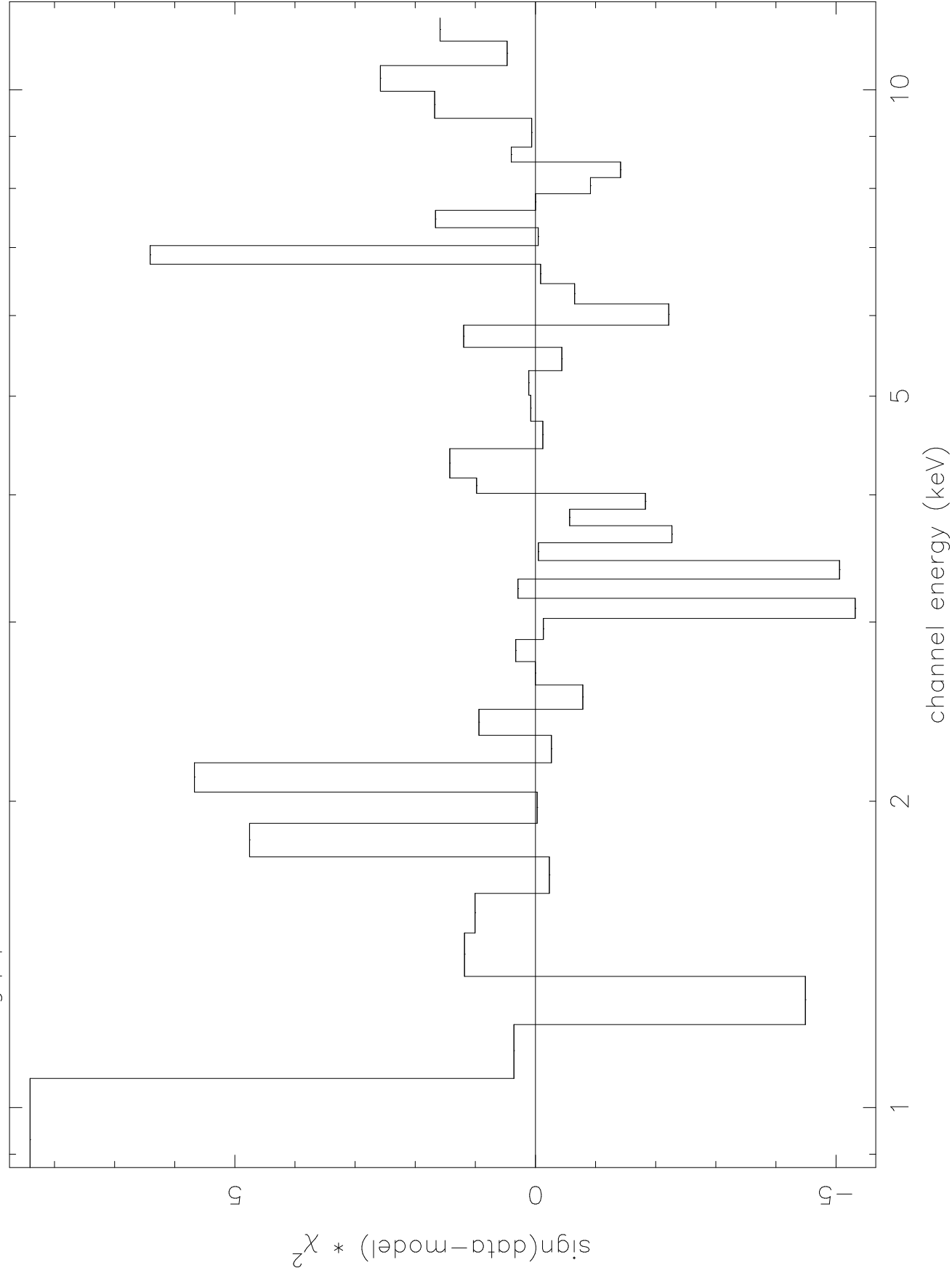